\documentclass[aps,prl,superscriptaddress,preprint]{revtex4-2}
\usepackage{graphicx}
\usepackage{hyperref}
\usepackage{amsmath}
\usepackage{color}
\usepackage{newtxtext,newtxmath}

\begin{document}

\title{\Large Predicting macroscopic properties of amorphous monolayer carbon via pair correlation function}

\author{Mouyang Cheng}
\thanks{Equal contribution}
\affiliation{School of Physics, Peking University, Beijing 100871, People's Republic of China}

\author{Chenyan Wang}
\thanks{Equal contribution}
\affiliation{School of Physics, Peking University, Beijing 100871, People's Republic of China}

\author{Chenxin Qin}
\thanks{Equal contribution}
\affiliation{School of Physics, Peking University, Beijing 100871, People's Republic of China}

\author{Yuxiang Zhang}
\thanks{Equal contribution}
\affiliation{School of Physics, Peking University, Beijing 100871, People's Republic of China}

\author{Qingyuan Zhang}
\affiliation{School of Physics, Peking University, Beijing 100871, People's Republic of China}

\author{Han Li}
\affiliation{School of Physics, Peking University, Beijing 100871, People's Republic of China}

\author{Ji Chen}
\email{ji.chen@pku.edu.cn}
\affiliation{School of Physics, Peking University, Beijing 100871, People's Republic of China}
\affiliation{Interdisciplinary Institute of Light-Element Quantum Materials and Research Center for Light-Element Advanced Materials, Peking University, Beijing 100871, People's Republic of China}
\affiliation{Frontiers Science Center for Nano-Optoelectronics, Peking University, Beijing 100871, People's Republic of China}

\begin{abstract}
Establishing the structure-property relationship in amorphous materials has been a long-term grand challenge due to the lack of a unified description of the degree of disorder.
In this work, we develop SPRamNet, a neural network based machine-learning pipeline that effectively predicts structure-property relationship of amorphous material via global descriptors.
Applying SPRamNet on the recently discovered amorphous monolayer carbon, we successfully predict the thermal and electronic properties.
More importantly, we reveal that a short range of pair correlation function can readily encode sufficiently rich information of the structure of amorphous material.
Utilizing powerful machine learning architectures, the encoded information can be decoded to reconstruct macroscopic properties involving many-body and long-range interactions.
Establishing this hidden relationship offers a unified description of the degree of disorder and 
eliminates the heavy burden of measuring atomic structure, opening a new avenue in studying amorphous materials.

\end{abstract}

\maketitle

Unravelling the connection between material structure and physical property is the foundation to discover new principles in condensed matter physics. 
Structure-property relationship (SPR) can be extremely subtle in disordered systems due to their complex and chaotic nature\cite{anderson1995through}. 
Two-dimensional (2D) amorphous materials appear as the most promising platform to tackle this challenge because of the possibility to visualize the structure at the atomic level in experiments via advanced transmission electron microscopy (TEM). \cite{huang2013imaging,hong2020ultralow,toh2020synthesis,joo2017realization,tian2023disorder,bai2024nitrogen}.  
Among 2D amorphous materials, 
amorphous monolayer carbon (AMC) stands out, providing critical opportunities to establish the complex SPR \cite{toh2020synthesis,tian2023disorder,felix2020mechanical,kim2023amorphous}.
On one hand, 
the degree of disorder of AMC can be controlled by growing conditions.
On the other hand, macroscopic properties such as electric conductivity and thermal transport properties can be measured and tuned in a wide range.

Early theoretical attempts focused on direct computational modelling of the atomic structure and the calculation of physical properties \cite{van2012insulating,antidormi2022emerging,cheng2023regulating,antidormi2020thermal,antidormi2022emerging,felix2020mechanical,fan2021linear}.
These studies offer important insights to the understanding of AMC, indicating SPR exists in an unconventional manner.
For instance, Tian et al. propose that the electric conductivity of AMC is linked simultaneously to two order parameters, namely the medium range order and the density of conducting sites \cite{tian2023disorder}.
However, to establish SPR of AMC, there are still a few major limitations to overcome. 
The first limitation arises from the inadequate sampling of atomic configurations of AMC. 
Theoretical works mostly assumed continuous random networks for AMC structure \cite{wright2014great}, but recent experimental works revealed large-scale voids and other complex defects are important structural features \cite{tian2023disorder}.
Secondly, accurate computations of properties such as electric conductivity and thermal conductivity
are rather time consuming and could not be applied to large-scale realistic amorphous materials.
Last but not the least, it is unclear how to find an optimal descriptor to encode the structural information.
Conventional physics studies would prefer simple scalar order parameters such as the variance of ring and bond \cite{van2012insulating}, the Steinhardt order parameters \cite{kotakoski2011point} and the medium range order parameter \cite{tian2023disorder}, which could lead to huge information loss.
Another idea is to adapt the brute-force approach widely used in the machine learning community, i.e. employing the full atomic structure as the input and utilizing the graph neural network to encode information directly \cite{gilmer2017neural,reiser2022graph}. 
However, such brute-force approaches would provide very limited physical insights to help the understanding of AMC.
In addition, measuring the atomic structure is in fact one of the most difficult tasks for amorphous materials including AMC, which again limits the usefulness of existing machine learning architectures.

\begin{figure*}[!htbp]
  \centering
  \includegraphics[width=0.98\textwidth]{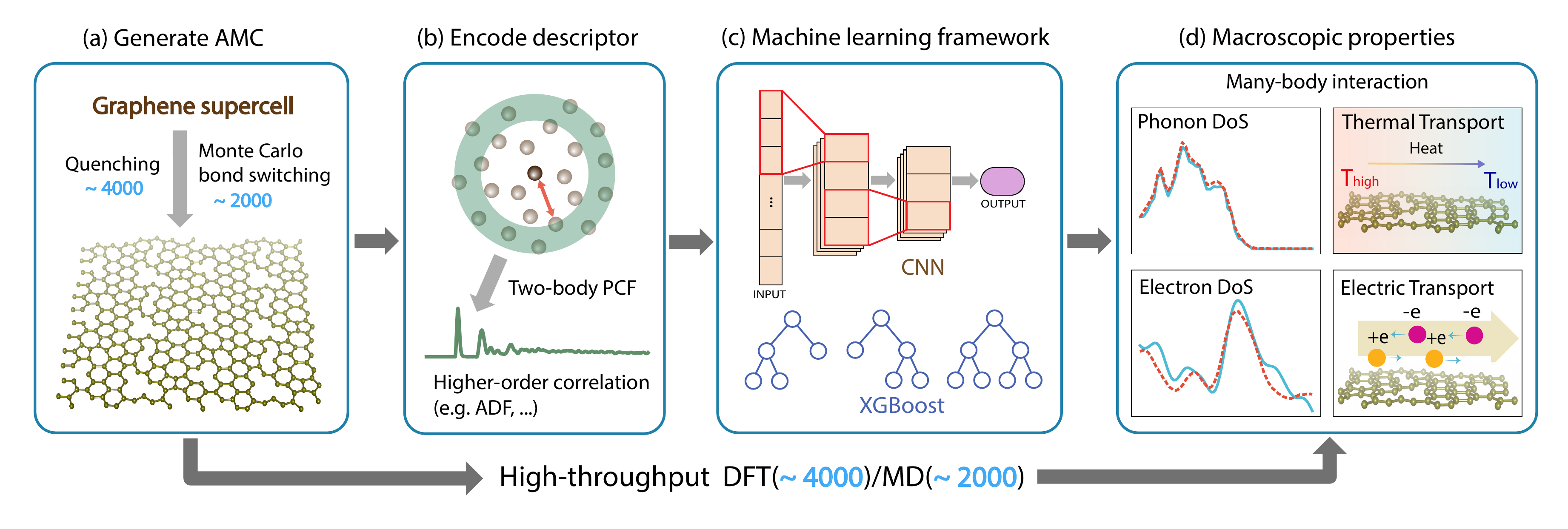} 
  \caption{\textbf{Schematic of the SPRamNet workflow for predicting electronic and thermal properties of amorphous monolayer carbon (AMC) from pair correlation function (PCF)}. 
  \textbf{(a)} AMC structures are generated by quenching or Monte Carlo bond switching algorithm from the pristine graphene supercell. The electronic and phonon properties for AMC are simulated via high-throughput density functional theory (DFT) and molecular dynamics (MD), respectively. 
  \textbf{(b)} The structural information is encoded using PCF highlighted as the green curve, while other higher-order correlation inputs are optional. \textbf{(c)} These global descriptors serve as the input to a machine learning pipeline based on convolutional neural network (CNN) and Extreme Gradient Boosting (XGBoost). 
  \textbf{(d)} Subsequently, the machine learning models can be trained to predict macroscopic electronic and thermal properties.
  }
  \label{fig1}
\end{figure*}

To break through all these limitations, in this work we introduce SPRamNet, standing for \textbf{S}tructure \textbf{P}roperty \textbf{R}elationship for \textbf{a}morphous \textbf{m}aterials with \textbf{Net}works, a machine-learning framework that links macroscopic electronic and thermal properties of amorphous systems (Fig. \ref{fig1}). 
Using AMC with varying degrees of disorder, we demonstrate the predictive power of SPRamNet. 
The overall goal is to train a model capable of predicting AMC spectral and transport properties based on physical descriptors as simple as possible, such as the pair correlation function (PCF).
PCF can be obtained when the atomic structure is determined, and it can also be measured experimentally without knowing the full atomic structure.
We train networks separately on density of states for electrons ($el$-DoS) \& phonons ($ph$-DoS), electrical conductivity and thermal conductivity. 


We generate 4,000 AMC configurations, each containing between 90 and 260 carbon atoms per supercell, for electronic property training. Additionally, we generate 2,000 configurations, each containing between 700 and 925 carbon atoms per supercell for phonon property training (Fig. \ref{fig1} (a)). Details on the number of atoms per unitcell are shown in Supplementary Information (SI, with additional references \cite{thompson2022lammps,lindsay2010optimized,fan2015force,boone2019heat,matsubara2020evaluation,kresse1996efficient,kresse1999ultrasoft,perdew1996generalized,wang2021vaspkit,pedregosa2011scikit,paszke2019pytorch}) II.
Starting from pristine graphene, we employ two typical techniques for creating diverse AMC structures: one is based on the Monte Carlo bond switching algorithm\cite{von2003realistic,wooten1985computer}, while the other is the quenching algorithm\cite{kumar2012amorphous}.
For the phonon dataset, all 2,000 structures are generated strictly in the two-dimensional plane, with half created using quenching algorithms and the other half using the Monte Carlo bond switching algorithm.
For the electron dataset, 2,000 structures are generated by two-dimensional quenching, while the other 2,000 structures are produced by bond switching in three-dimensional space to allow the natural out-of-plane oscillation. 
Subsequently, we perform high-throughput density functional theory (DFT) calculations to obtain $el$-DoS and electric conductivity. 
$ph$-DoS and thermal conductivity are calculated using classical molecular dynamics (MD)
with the Tersoff\cite{tersoff1988empirical} force field (Fig. \ref{fig1} (d)).
More details about data preparation can be found in SI.I.

The key of our workflow is to construct an appropriate descriptor to encode the structural information (Fig. \ref{fig1} (b)).
A suitable descriptor should both be scalable -- applicable for arbitrary sizes of input supercells -- and allow E3 \& permutation-invariant outputs, meaning the neural network prediction remains unchanged under atomic permutations and E3 transformations such as rotation, translation, and inversion.
As mentioned above, a brute-force approach is to build a message passing graph neural network (MPNN)\cite{gilmer2017neural,reiser2022graph} with multiple localized descriptors on each nodal atoms. 
Scalable and permutation-invariant output features can thus be guaranteed via global aggregation.
However, the usefulness of localized descriptors heavily depends on the availability of atomic configurations. 
In SPRamNET we choose global descriptors such as PCF and the angular distribution function (ADF) to encode structural information of AMC.
In particular, PCF describes how density varies as a function of distance from a reference particle, and is of special physical significance, as it is directly related to the system's structure factor via inverse Fourier transform.
Pair correlation function can be obtained in experiments using X-ray or neutron diffraction\cite{dinnebier2008powder}, bypassing quite involved measurements of atomic configurations, hence choosing it as a descriptor overcomes the greatest challenge in experimental studies of amorphous materials.
Moreover, it is straightforward to check that PCF ensures both scalability and invariance while simplifying the computational complexity, making it an efficient and physically insightful choice for modelling AMC structures. 

With PCF readily calculated from structural input, in SPRamNet
we discretize PCF and target physical properties as fixed-size input and output of the network, respectively (Fig. \ref{fig1} (c)). 
In principle, the network can be subsequently trained to reconstruct macroscopic properties of AMC.
For different property prediction tasks, we make minor adjustments to the network dimensions to make them compatible to the outputs, either vectorized $el$-DoS or $ph$-DoS, or simply scalar conductivity values $\sigma$ or $\kappa$. 
One can also truncate PCF at different distance to further simplify the input information.

\begin{figure}[!htbp]
  \centering
  \includegraphics[width=0.7\textwidth]{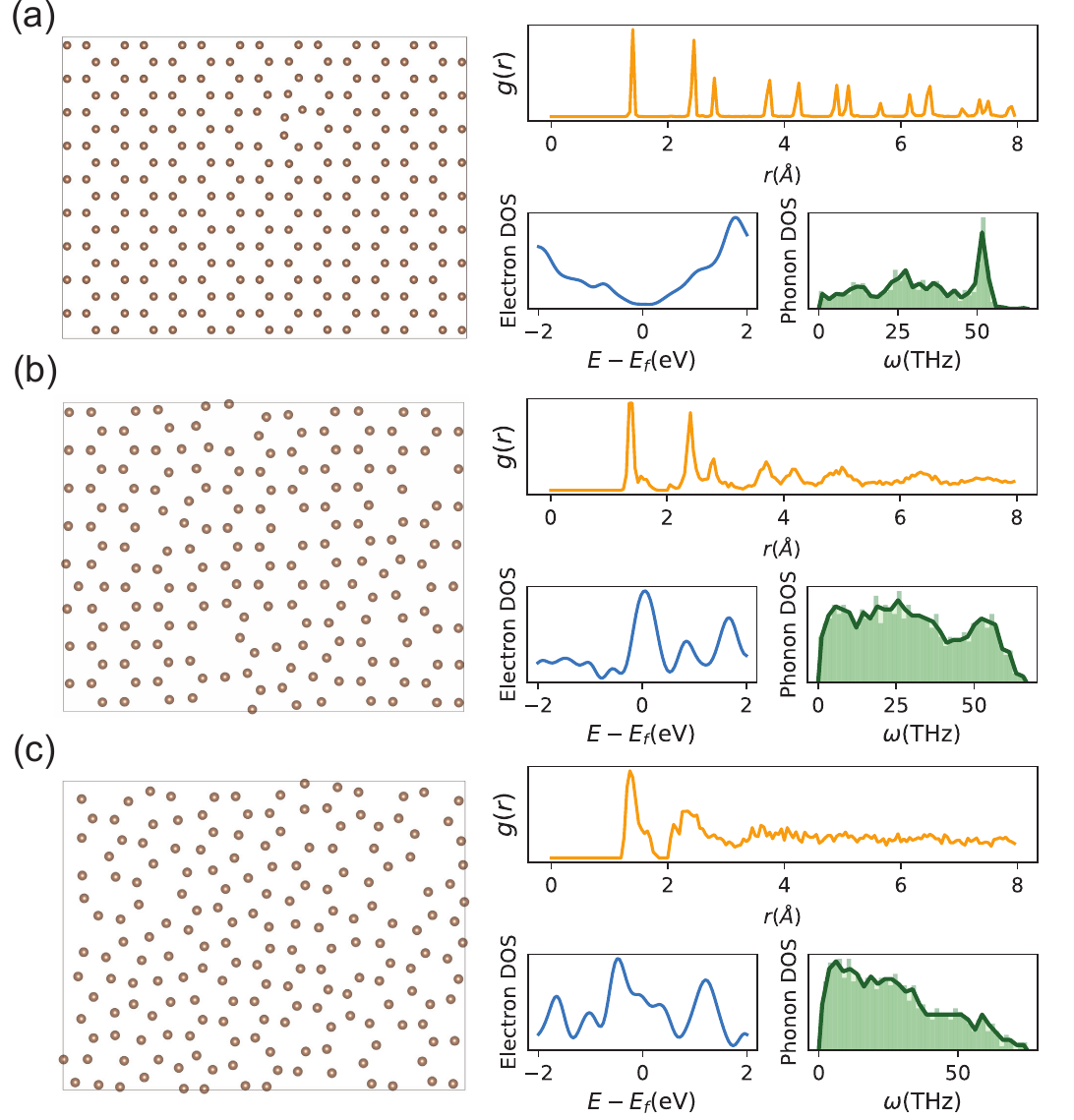}
  \caption{\textbf{Display of dataset generated for SPRamNet training.} Three typical types of AMC structures with different degrees of disorder: \textbf{(a)} graphene structure with point defects; \textbf{(b)} near-pristine graphene structure with minor perturbation and defects; and \textbf{(c)} highly amorphous carbon structure, along with their corresponding PCF $g(r)$, density of states ($el$-DoS), and phonon density of states ($ph$-DoS):}
  \label{fig2}
\end{figure}
Before presenting the machine learning results, let's have a deeper analysis of the AMC structures in the dataset, which covers a wide range of degree of disorder. 
Fig. \ref{fig2} (a)-(c) show three AMC structures with increasing degree of disorder and their corresponding spectral properties.
We set the frequency range for $ph$-DoS within $f\in[0, 100]$ THz, and the energy scale for $el$-DoS within $E\in[-2, 2]$ eV w.r.t the Fermi level $E_f$, since it covers the energy window where most important physical processes occur. 
PCF is shown within a cutoff radius of $r_{\text{max}}=8 \rm{\mathring{A}}$.
We note that the PCF descriptor $g(r)$ should be normalized w.r.t the radial distance $r$ and number of atoms $N$, so that it becomes scalable and converges for increasing $r$. Also, additional care is required for a properly normalized $g(r)$ for quasi-2D structures in the electron dataset.
More details on normalizing the $g(r)$ is shown in SI.I.4.

As expected, the degree of disorder has substantial impacts on spectral properties of AMC. For the structure with minor defects (Fig. \ref{fig2} (a)), long range crystalline order is clearly present in $g(r)$, where multiple sharp peaks persist for $r>4 \rm{\mathring{A}}$. The $el$-DoS is near zero at the Fermi level, which also resembles the pristine graphene behaviour, and a sharp peak is observed near $f=48$ THz, corresponding to the G peak frequency and vibration mode in graphene.
As disorder gradually dominates in the AMC structure (from Fig. \ref{fig2} (a) to (c)), peaks for $r>4 \rm{\mathring{A}}$ become suppressed and gradually vanish, indicates absence of long range order in strongly disordered AMC structures.
Such loss of long range order induces great change in measurable physical properties.
This change is evident in $el$-DoS, where a centered peak appears at the Fermi level and then smears out as degree of disorder increases. Additionally, while the peak positions in $ph$-DoS remain largely unchanged, the heights and widths of these peaks are significantly altered, indicating dominance of other vibrational modes in AMC as disorder starts to play a important role. Apart from these spectral properties, transport properties also vary in three orders of magnitude within our dataset, which will be shown below and in SI.II.

\begin{figure}[!htbp]
  \centering
  \includegraphics[width=0.7\textwidth]{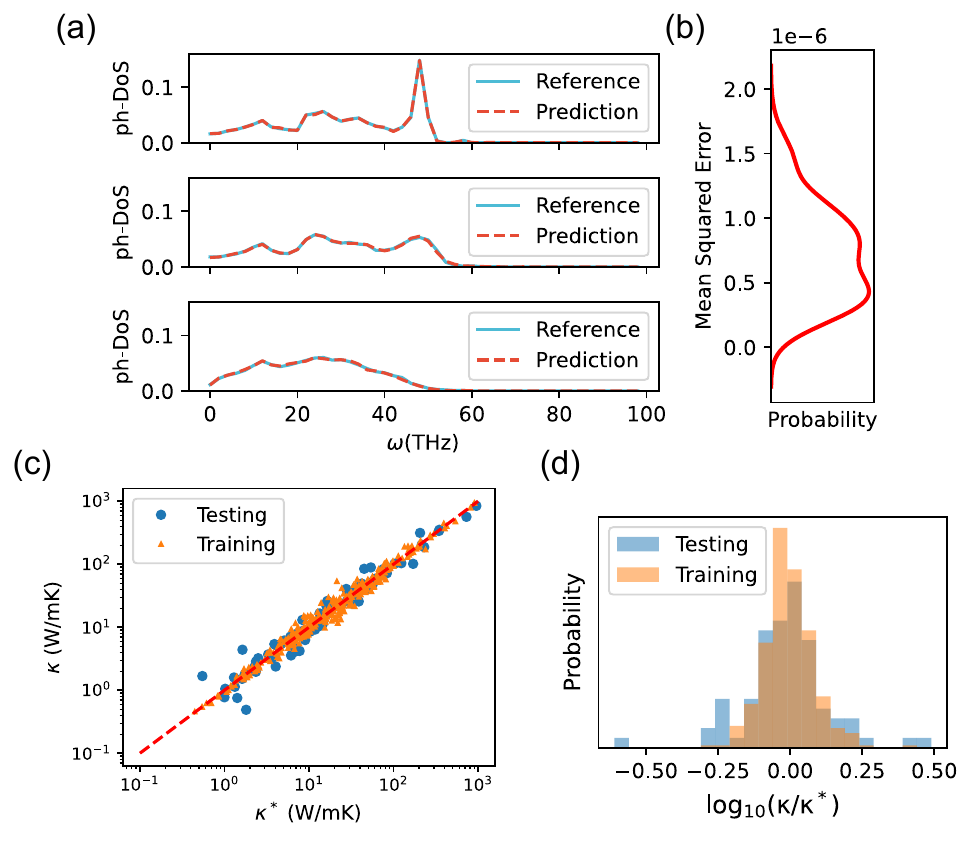}
  \caption{\textbf{SPRamNET Prediction of AMC thermal properties.} \textbf{(a)} Comparison between the calculated and the predicted $ph$-DoS in the testing set. \textbf{(b)} Distribution of MSE loss for testing results.
  \textbf{(c)} Predicted thermal conductivity $\kappa$ at $T=300 \rm{K}$ plotted against  
  the calculated reference ($\kappa^*$) on the logarithmic axes for comparison.
  \textbf{(d)} Distribution of log loss: $\rm{log}_{10}(\kappa/\kappa^*)$ for training and testing results.}
  \label{fig3}
\end{figure}

Evidence above hints the possibility of a structure property relation between PCF descriptor and electronic \& thermal properties, but to what extent can PCF reconstruct these macroscopic properties is yet to explore.
Fig. \ref{fig3} presents the results on thermal properties of AMC using a convolutional neural network.
The network architecture is based on a variant of LeNet\cite{lecun1998gradient} with optional residual connection\cite{he2016deep}. More details on the architecture and hyperparameter settings are shown in SI.III.1.
Fig. \ref{fig3} (a) displays the comparison of $ph$-DoS for three typical structures with different levels of degree of disorder, and we observe a prefect agreement between MD and machine learning results.
More quantitatively, the expectation of mean squared error (MSE) loss is as low as $\sim 10^{-6}$ in Fig. \ref{fig3} (b), which is negligible compared to the $ph$-DoS signal.
To further test whether PCF contains sufficient information as the structural input, we incorporate the three-body correlation function, the angular distribution function (ADF), as additional input (Fig. S8).
We find that such addition gives $\sim 3\%$ improvement in terms of MSE loss, indicating a marginal contribution of higher order correlation functions.

Similarly, SPRamNet shows good performance on the thermal conductivity.
Reference value of thermal conductivity $\kappa$ is calculated using MD at 300 K
employing the Green-Kubo (GK) formalism.
In Fig. \ref{fig3} (c), the predicted thermal conductivity ($\kappa$) is plotted against the reference value ($\kappa^*$) for both the training and the testing data.
In Fig. \ref{fig3} (d), the error distributions are plotted.
Overall, these findings above suggest that SPRamNet receiving PCF as descriptor alone can almost fully predict the thermal properties of AMC.
One can show the existence of SPR for model systems with pair-wise interaction \cite{chandler1987introduction}, but the success of SPRamNet reveals that more SPRs are hidden in realistic materials containing interactions beyond pair-wise types.

\begin{figure}[!htbp]
  \centering
  \includegraphics[width=0.7\textwidth]{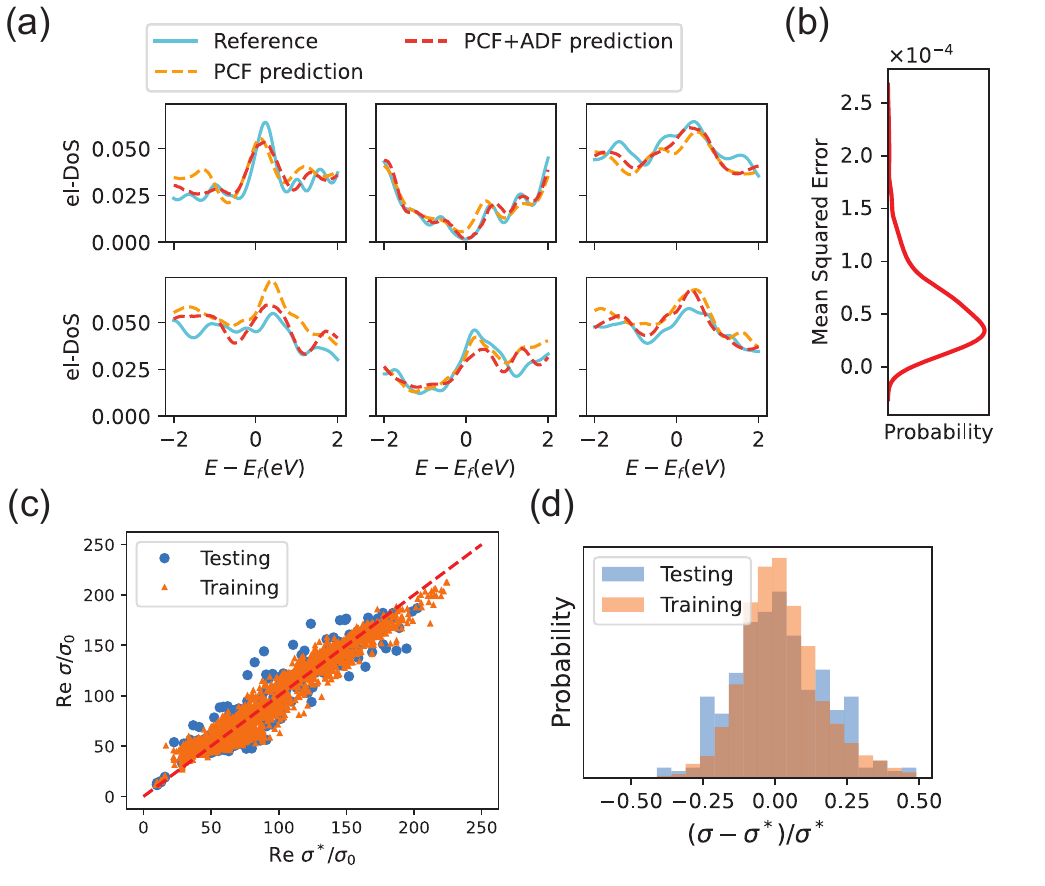}
  \caption{\textbf{SPRamNet prediction of electronic properties.} \textbf{(a)} Examples comparing the calculated $el$-DoS with the predicted $el$-DoS in the testing set, for both PCF and PCF+ADF prediction. $\sigma_0 = 1$ kS/m. \textbf{(b)} Distribution of MSE loss for PCF+ADF testing results.
  \textbf{(c)} Predictions of diagonal real electric conductivity Re $\sigma$ at zero temperature. Predicted $\sigma$ are plotted against DFT calculated reference $\sigma^*$ for comparison. \textbf{(d)} Distribution of relative error $(\sigma-\sigma^*)/\sigma^*$.}
  \label{fig4}
\end{figure}

Moving to electronic properties, the model training immediately becomes more difficult with the same network used for thermal properties, indicating the SPR for electronic structure involves more degrees of freedom in AMC structure as expected.
For $el$-DoS, we find it is necessary to improve the depth of the network to improve the expressiveness. 
As shown in Fig. \ref{fig4} (a), a deeper network can readily improve the prediction (red and orange curves) of $el$-DoS to a satisfactory level compared with the DFT results (blue curves).
The expected MAE loss is lower than $5\times 10^{-5}$ in Fig. \ref{fig4} (b), and SPRamNET can still well reconstruct the main features of DFT calculated $el$-DoS.
Going through all the data, one can still notice visible deviations for some AMC structures. 
With the inclusion of ADF in the input, the predicted $el$-DoS can be further improved.
From the MSE loss, the improvement is estimated $\sim 10\%$, which further supports that PCF contains sufficient information to qualitatively reconstruct electronic properties (SI.III.2).

For the electric conductivity $\sigma$, 
we take the frequency limit $\omega \to 0$ for the real part of optical conductivity $\sigma(\omega)$ calculated using DFT.
In this more challenging case, we find the accuracy of convolutional network can only be mildly improved via enlarging the network depth. 
To this end, an alternative machine learning architecture, namely the Extreme Gradient Boosting (XGBoost)\cite{chen2016xgboost} algorithm, is implemented and reaches the highest accuracy among all the settings tested (SI.III.3).
XGBoost is more suitable and easier to tune for scalar predictions in complex tasks, hence it is integrated as a part of SPRamNet.
The predicted electric conductivity are shown in Fig. \ref{fig4} (c), where both training and testing results are displayed for comparison. 
The relative testing error shown in Fig. \ref{fig4} (d) is mostly less than 25\%, while the root mean square error (RMSE) of the testing result is 13.2$\sigma_0$ for the diagonal conductivity.
The results suggest that with a suitable machine learning architecture, PCF can still be used as an effective descriptor for electrical conductivity.

Although the performance of SPRamNet is slightly worse for electronic properties compared to thermal properties, the achievement is a bigger surprise.
We also note that predicting electronic properties such as the electronic density of states ($el$-DoS) remains a significant challenge in the literature. Compared to previous works focusing on $el$-DoS prediction for crystals using graph network embedding\cite{louis2020graph,kong2022density}, our SPRamNet demonstrates superior performance using a much more simplified global descriptor PCF, particularly in accurately capturing the key peaks in the $el$-DoS. These peaks are critical for extracting material behavior, such as thermoelectric and optical properties.

PCF carries structural information on amorphous material by encoding the atomic pairs at various distances. 
The descriptor itself is a two-body metric, but many-body information are indirectly accounted.
This hints at a potential reduction in the effective dimensionality of the problem, where two-body interactions encapsulate much of the behavior that would traditionally be attributed to higher-order interactions.
Moreover, the success of SPRamNet suggests that simple descriptors may contain rich information, which were not fully explored in traditional physics studies.
To test the limit of a minimal descriptor, we analyze the effective range of $g(r)$.
The results reveal that all significant contributions are confined to features in $g(r)$ with radial distances $r<1.8\rm{\mathring{A}}$ (SI.III.4). 
This indicates that a very small range of $g(r)$ could already reflect the most relevant structural information.

To summarize, we propose SPRamNet, a machine learning workflow to reconstruct both electronic and phonon properties of AMC by using structure correlation functions as global descriptors.
The SPRamNet framework establishes the complex SPR of AMC, and demonstrates that AMC's macroscopic properties can be captured through the simplest pairwise correlations.
Leveraging the power of machine learning, our results indicate that even as a global descriptor representing the average effects of local environment, PCF is still generally linked to electronic, phonon, and potentially other measurable properties even for many-body interactions.
SPRamNet could be generalized to suit other glassy systems beyond two-dimension and single carbon element, such as amorphous alloy and semiconducting compounds.
In existing machine learning studies of disordered systems, local structure features are often considered as the main input and their aggregations are used to characterize the degree of disorder \cite{chapman2023quantifying,aykol2023predicting}.
In contrast, this study promotes the use of simple yet physically meaningful descriptors for machine learning studies of complex amorphous materials, which offers several other advantages, including facile training and high interpretability. 
Finally, because SPRamNet only takes PCF as the input, it can take the advantage of many mature experimental techniques for amorphous materials.
At the current stage, experimentally measuring atomic coordinates of amorphous materials is still one of the greatest challenges in the field. 
Local structure features would require the information of atomic coordinates, but for a globally averaged PCF one can circumvent the issue by employing diffraction techniques such as X-ray and neutron scattering, which are much more available in experimental labs.

\section{Acknowledgements}
This work is supported by National Key R\&D Program of China under Grant No. 2021YFA1400500, the Strategic Priority Research Program of the Chinese Academy of Sciences under Grant
No. XDB33000000, the National Natural Science Foundation of China under Grant No.
12334003, and the Beijing Municipal Natural Science Foundation under Grant No. JQ22001. The authors are grateful for computational resources provided by the High Performance Computing Platform of Peking University. 

\bibliography{refs_introduction.bib,refs_method.bib,refs.bib} 
\end{document}